\documentclass[epj]{svjour}
\usepackage{graphics}
\usepackage{graphics}
\usepackage[pdftex]{graphicx}
\usepackage{xspace,colortbl}
\usepackage{amsmath}
\usepackage{amssymb}
\usepackage[T1]{fontenc}
\usepackage{float}
\usepackage[utf8]{inputenc}
\usepackage{authblk}
\usepackage{graphicx}
\usepackage[T1]{fontenc}
\begin{document}
\title{Evolution of spherical overdensity in Chaplygin gas model  }
\author{Amin Rezaei Akbarieh\inst{1}\thanks{am.rezaei@tabrizu.ac.ir}\and Mohammad Ahmadi\inst{1}\thanks{mohammad.ahmadii7373@gmail.com} \and Yousef Izadi\inst{2}\thanks{yousef\_izadi@uml.edu} \and Shahabeddin M. Aslmarand\inst{3}\thanks{smostafanazh2016@fau.edu} \and Warner A. Miller\inst{3}
\thanks{wam@fau.edu}%
}                     
%
%
\institute{Department of Theoretical Physics and Astrophysics, University of Tabriz, 51666-16471, Tabriz, Iran \and Department of Physics and Applied Physics, University of Massachusetts, Lowell, MA 01854, USA \and
Department of Physics, Florida Atlantic University, Boca Raton, FL 33431, USA}
\date{Received: date / Revised version: date}
%
\abstract{ Even though many scalar field models
           of dark energy have been considered in the literature, there is another interesting class
           of dark energy models involving a fluid known as a Chaplygin
           gas. In addition to describing the dark energy, both scalar-tensor model and the Chaplygin gas model are suitable candidates for
           explaining the spherical cosmological collapse. One of the most well-known scalar field models is the quintessence
           model, which was first introduced to explain an accelerating expanding universe. Using a special form of the quintessence model that is equivalent to Chaplygin gas, we describe evolution of a spherical collapse. We study the cosmological properties of the quintessence field with a special potential. In addition to the quintessence model, that can be converted into a Chaplygin gas model in a particular case, we claim that the fixed-potential tachyonic model is equivalent to the Chaplygin gas model. In this work, we obtain the spherical collapse parameters: the virialized overdensity parameters, radius, the energy density at the turnaround moment, etc. We compare the results of the proposed model with the standard model of cosmology and the Einstein–de Sitter model. We show that the formation of the large-scale structures within the framework of a Chaplygin gas model happens earlier than predicted in the standard model.
\PACS{
      {PACS-key}{discribing text of that key}   \and
      {PACS-key}{discribing text of that key}
     } 
} 
\maketitle
\section{Introduction}
\label{intro}
Almost decades after the introduction of the Einstein-de sitter dust model, observations of the redshift in type Ia supernova suggested a positive acceleration of the universe \cite{Riess:1998cb}. These results led to the proposal of an unknown exotic matter which is called dark energy and it satisfies the condition $\rho +3P <0$ where $\rho$ represents the density of the matter and $P$ denotes the pressure. There have been many candidates for the dark energy, the simplest model is based on the cosmological constant which is not dynamic. The cosmological constant faces two issues, "fine tuning" and "the coincidence problem" \cite{Kofinas:2017gfv}. Models of interacting dark energy were introduced to solve the problem of coincidence. Quintessence \cite{Carroll:1998zi} and K-essence \cite{GonzalezDiaz:2003rf} models were proposed to solve such issues, these models are based on either non-minimally or minimally coupled scalar fields \cite{Perenon:2016itr,Bhattacharya:2015wlz}.
Although these models face the challenge of fine-tuning, which states that, in terms of mathematical laws and probabilities, the universe could have been created in innumerable different ways. However, our universe is only a very special case. In recent literature some models have been investigated in which, in order to solve the fine tuning problem, it is necessary to consider dark energy in the initial stage of the universe \cite{Riess:1998cb}. Therefore, to explain the number of structures at present, the contribution of dark energy cannot be ignored. On the one hand, the process of structure formation should occur earlier and much slower than predicted in the standard cosmological model. Additionally, the analytical calculations show that by considering some assumptions for the collapse of a homogeneous and isotropic sphere, one can conclude that the number of structures formed within the scalar-tensor models is more than predicted in the $\Lambda CDM$ model. Also, numerical analyses show that the classical spherical model is in good agreement with the simulations \cite{Pace:2010sn}. The dynamics of the overdensity parameter depend on the evolution of the flux, which causes the universe to expand. So, dark energy models, such as scalar-tensor models, can describe the growth of structures.\\
One of the most interesting ideas for the unification of dark matter and dark energy is the Chaplygin gas model in which the dark sector of the universe is expressed in terms of a single component that serves for both dark energy and dark matter. Chaplygin gas is a perfect fluid that behaves like a pressureless fluid in the early universe while it causes the accelerated expansion of the universe at the late time. It has been shown that the resulting evolution of the universe is compatible with the observational results \cite{Kamenshchik:2001cp}. The model states that the value of the effective cosmological constant is increasing. The simplest model of the Chaplygin gas satisfies the following equation of state
\begin{equation}
\label{eqn:Chap}
P=-\frac{A}{\rho},
\end{equation}
in which $A$ is a constant. In \cite{Bento:2002ps}, the scenario based on the dynamics of a generalized D-brane in a space-time $(D + 1, 1)$ is investigated. They stated the conditions for homogeneity and demonstrated that the equation of state describes the evolution of a universe from a matter-dominated phase to the cosmological constant dominated phase. They wrote the effective equation of state as follows
\begin{equation}
\label{eqn:Chap1}
P=-\frac{A}{\rho^{\alpha}},
\end{equation}
which is the state equation of the generalized Chaplygin gas with $0<\alpha\leq1$. In \cite{Bento:2002yx}, the constraints on this model of cosmic background radiation are discussed. In this paper, from the results of Archeops for the location of the first peak and, from the results of BOOMERANG for the location of the third peak, also, from the supernova, high-redshift observation, and gravitational lensing statistics \cite{Dev:2002qa}, it is shown that this model is completely distinguishable from the standard model of cosmology. In line with previous works in \cite{Amendola:2003bz}, the WMAP temperature power spectrum and supernova data are compared with the generalized Chaplygin gas model, and it has been shown that the parameter $\alpha$ is in the range $(0,0.2)$ with $95\%$ confidence. The authors have also shown that the generalized model at state $\alpha = 1$ was ruled out as a candidate for dark energy with an accuracy of more than $99.99\%$. Furthermore, by examining this model in the non-flat background, one can see that the non-flat state corresponds to a flat case with an accuracy of $68\%$ \cite{Bertolami:2004ic}. In \cite{Debnath:2004cd,Benaoum:2002zs} the modified Chaplygin gas model with the equation of state $P=A\rho-B/\rho^\alpha$ in which $A$ and $B$ are constant, is considered. Assuming that the equation of state for the modified model of the Chaplygin gas is valid from the radiation period (A = 1/3, and for very large densities) to the current time (small densities), they show that their model can describe the accelerated expansion of the universe. In \cite{Bilic:2001cg}, it is shown that the inhomogeneous model of Chaplygin gas can explain dark matter and dark energy together in a geometric setting reminiscent of M-theory. Since the scalar field can explain both the holographic dark energy and the Chaplygin gas model, the relationship between the (interacting) holographic dark energy density and the energy density of the Chaplygin gas model is studied in \cite{Setare:2007jw,Setare:2007mp}.\\
Models that describe the dark energy issue not only explain the accelerated expansion of the universe, but they also play a fundamental role in the formation of cosmological structures. The large scale structures we see today are formed by the small perturbations in the inflationary phase of the early universe, and under the influence of gravity, these perturbations grow. In 1972 Gunn and Gott proposed the spherical collapse model \cite{Gunn:1972sv}, which led them to reach a simple explanation for the growth of the over dense structures. To understand the formation of the large scale structures and the growth of perturbations in a matter-dominated universe within the spherical collapse model, one can consider a spherical region with a radius that expands in time and with a non-uniform density. According to Birkhoff's theorem \cite{Birkhoff}, the evolution of this radius depends only on the limited mass within the region. With the expansion of the cosmic background, this perturbed region begins to expand, and its radius reaches a maximum value. However, after a while, depending on its mass and due to its high density, it decouples from the Hubble background and starts evolving in a reverse process that decreases the radius or collapse in the non-linear region. Note that the collapse of this region does not proceed to a singular point. According to the virial theorem, the collapse of these perturbations ceases at half of the maximum radius, which leads to the formation of the structure.
The spherical symmetry makes the problem easier to solve. Therefore, the spherical collapse model is a suitable model for understanding the evolution of the perturbations. Additionally, this model is successful in reproducing the simulation results \cite{Pace:2010sn,Pace:2013pea,Hiotelis:2012ff,DelPopolo:2006au}.\\
 The studies and progresses mentioned in scalar-tensor theories were within the framework of Einstein's standard theory of general relativity. In 2004, Motta and Brock investigated the minimally coupled quintessence model under different potentials \cite{Mota:2004pa} and they concluded, in models with the assumption of non-minimal coupling of the scalar field, the evolution of the spherical overdensity parameter is quite different from the minimally coupled model. The non-linear evolution of structures within the framework of non-minimally coupled  quintessence models was studied by Pace et al \cite{Pace:2010sn}. They conclude that the value of the virial overdensity and the linear density threshold for the spherical collapse are very close to the predictions of the standard cosmological model. Note that these two parameters play an essential role in the spherical collapse model. In the paper by Fan et al \cite{Fan:2015rha}, the non-minimally coupled model in the metric and Palatini formalisms is studied. It is shown that metric and Palatini formalisms lead to different results for linear growth rates. However, this difference is so small, and it is not observable now.\\
Therefore, in this paper we study the non-linear evolution of structures in the framework of the spherical collapse model using the Chaplygin gas model. Section \ref{sec:1} presents an introduction to the quintessence model, and shows that by choosing an appropriate potential for the quintessence, one obtains the equation of state of the Chaplygin gas. In section \ref{sec:2}, this article studies the non-linear evolution of structures and shows that the quintessence model, in a special form, is equivalent to the Chaplygin gas model and plays an effective role in the spherical collapse model. We also discuss that the results of this model are the possibility for the formation of the structures in this framework occurs sooner than the cosmological standard model. Section \ref{sec:3} discusses that the tachyon field under a constant potential is equivalent to the Chaplygin gas model and examines the parameters of the spherical collapse model within this framework. Also, we show that the formation of structures under the Chaplygin gas model happens faster than what the standard cosmological model predicts. Section \ref{sec:4} contains a conclusion to this article.
\section{Quintessence field as a Chaplygin gas}
\label{sec:1}
\begin{figure}
\centering
\includegraphics[width=0.8\linewidth]{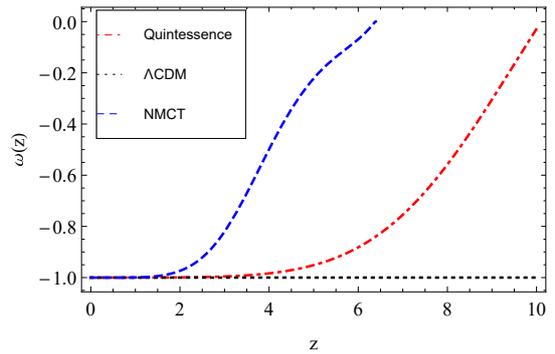}
\caption{The behavior of EoS parameter $\omega(z)$ versus the redshift for the $\Lambda CDM$, quintessence and  non-minimally coupled tachyon models (NMCT).}
\label{fig1}
\end{figure}

\begin{figure}
\centering
\includegraphics[width=0.8\linewidth]{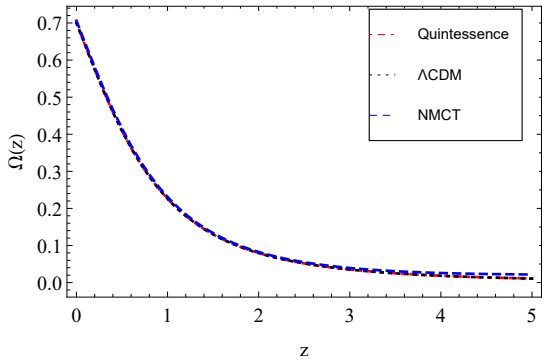}
\caption{The energy density versus the redshift for the $\Lambda CDM$, quintessence and NMCT models .}
\label{fig2}
\end{figure}

In 1988, Ratra and Peebles \cite{Ratra:1987rm} proposed a scenario in which one can use a scalar field which is universal, rolling, self-interacting, and homogeneous to describe the dynamical energy density of the universe. Models based on this scenario are known as quintessence. Also, the quintessence fields are one of the best candidates for the scalar field to explain the formation of the large scale structures in the universe. Following Ratra and Peebles, quintessence fields were studied more by Robert R. Caldwel, Duhal Dave and Paul Steinhardt \cite{Caldwell:1997ii}. Some people even considered it as the fifth fundamental force of nature. The action for the quintessence field can be written as
\begin{equation}
S=\int dx^4 \sqrt{-g} \Bigg{[} -\frac{1}{2}g^{\mu\nu}\partial_\mu \phi \partial_\nu \phi-V(\phi)\Bigg{]},
\end{equation}
where $V(\phi)$ is the scalar field potential. We assume that the field is only a function of time $\phi=\phi(t)$. We use the FLRW metric to describe the expanding universe, so we have $\sqrt{-g}=a(t)^3$ where $a(t)$ is the scale factor. Then the point-like Lagrangian of this field becomes
\begin{equation}
\mathcal{L}=a^3\Bigg{[}\frac{1}{2} \dot{\phi}^2-V(\phi)\Bigg{]}.
\end{equation}
One can easily write the equation of motion as
\begin{equation}
\ddot{\phi}+3H\dot{\phi}+V'(\phi)=0.
\end{equation}
If we choose the potential in this model as follows
\begin{equation}
\label{eqPot}
V(\phi)=\frac{\sqrt{A}}{2}\bigg{(} \cosh(\sqrt{3}\kappa\phi)+\frac{1}{\cosh(\sqrt{3}\kappa\phi)}\bigg{)},
\end{equation}
one can conclude that the scalar quintessence model is equivalent to the Chaplygin gas model (Appendix A). For the aforementioned potential, $A$ and $\kappa$ are constants. Many observations place constraints on the model parameters, $\kappa$ and  $A$  should approximately take the values $k\simeq 0.06$, 0.28< A <0.550. The energy density and the pressure are obtained as
\begin{equation}
\rho_\phi=\frac{1}{2}\dot{\phi}^2+V(\phi),
\label{7}
\end{equation}
\begin{equation}
\label{8}
 P_\phi=\frac{1}{2}\dot{\phi}^2-V(\phi).
\end{equation}
One can use the expressions in (\ref{7}) and (\ref{8}) to find the equation of state parameter (EoS) as
\begin{equation}
\omega_\phi =\frac{P_\phi}{\rho_\phi}=\frac{\dot{\phi}^2-2V(\phi)}{\dot{\phi}^2+2V(\phi)}.
\end{equation}
In the slow-roll approximation $\dot{\phi}^2 << V(\phi)$, we have $P\simeq -\rho$. In Fig.\ref{fig1}, we draw a graph of the EoS parameter for the quintessence verse the redshift, and we compare it with the results of the $\Lambda CDM$ and non-minimally coupled tachyon models (NMCT).
As we see from the figure, $\omega$ is a constant $(\omega=-1)$ for the $\Lambda CDM$ model. However, for the quintessence and the NMCT models, at high redshifts the value of $\omega$ is slightly different from the standard model, while at low redshifts, it takes a value which completely agrees with the standard model.
One can use the components of the stress tensor to find energy densities of the field and matter as
\begin{equation}
\label{eq1}
\Omega_{\phi}=\frac{\rho_{\phi}}{3M_{pl}^2 H^2},
\end{equation}
\begin{equation}
\Omega_{m}=\frac{\rho_m}{3M_{pl}^2H^2},
\end{equation}
where $H$ and $M_{pl}$ are Hubble parameter and Planck mass respectively. The energy densities of the scalar field and matter are give by
\begin{equation}
\rho_{\phi}=\rho_{\phi0}a^{-3(1+\Omega_{\phi})},
\end{equation}
\begin{equation}
\rho_{m}=\rho_{m0}a^{-3},
\end{equation}
which leads to the continuity equations
\begin{equation}
\dot{\rho}_{\phi}+3H\rho_{\phi}(1+\Omega_{\phi})=0,
\end{equation}
\begin{equation}
\dot{\rho_{m}}+3H\rho_{m}=0.
\end{equation}
By taking the time derivative of equation \eqref{eq1}, we find the variations in the field density as
\begin{equation}
\dot{\Omega_{\phi}}=-\Omega_{\phi}H\Bigg{[} 3(1+\omega_{\phi})+2\frac{\dot H}{H}\Bigg{]},
\end{equation}
where
\begin{equation}
2\frac{\dot H}{H}= -3(1+\omega_{\phi}\Omega_{\phi}).
\end{equation}
Using the following change of variable
\begin{equation}
\frac{d}{d(\ln a)}=-(1+z)\frac {d}{dz},
\end{equation}
one can write the change of field density with respect to the redshift
\begin{equation}
\Omega _{\phi}^{\prime}=-3\Omega_{\phi}(\Omega_{\phi}-1)(\frac{(1+z)^2\phi^{\prime 2}-2V(\phi)}{(1+z)^2\phi^{\prime 2}+2V(\phi)})(1+z)^{-1}.
\end{equation}
Note that prime denotes the derivative with respect to $z$. The value of $\Omega _{\phi}$ approaches  to zero for the large redshifts (early times), while from the observations we know that the value of this parameter for small $z$ (present time) approximately equals to $0.7$. From Fig. \ref{fig2}, one can see that the value of $\Omega _{\phi}$ in the quintessence model is very close to the prediction of the standard model of cosmology.
Now, using the above equations, we find the relation between the Hubble parameter and the energy density of the scalar field as follows
\begin{equation}
E^{\prime}(z)=\frac{3}{2}\frac{E(z)}{1+z}(1+\omega_{\phi}\Omega_{\phi}),
\end{equation}
where
\begin{equation}
E(z)=\frac{H}{H_0},
\end{equation}
\begin{figure}
\centering
\includegraphics[width=0.8\linewidth]{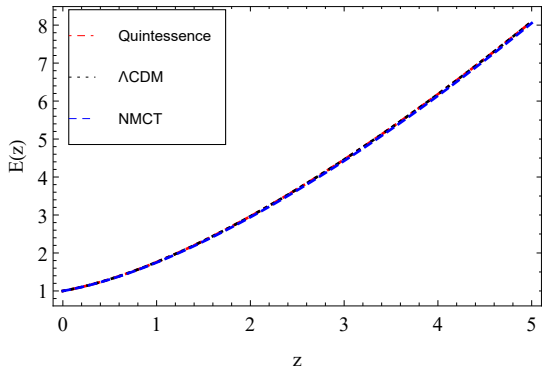}
\caption{The variations of the dimensionless Hubble parameter with respect to the redshift for the quintessence, minimally-coupled tachyon and the standard cosmological models.}
\label{fig3}
\end{figure}
in which $H_0$ is the Hubble parameter at the present time. One can see that for $z=0$, we have $E(0)=1$. Formation of large scale cosmological structures such as super clusters is one of the main questions in cosmology and astrophysics.\\
Probably some factors such as inhomogeneous distribution of matter are responsible for the growth of  the perturbations in the early universe that caused the formation of the cosmological structures during the period of matter-dominant. To understand the perturbations growth in the matter-dominant epoch which formed the structures that we have today, we consider a slice of the expanding universe with a radius of $R$ and non-uniform density $\rho = \bar{\rho} (1 + \delta)$ (where $\delta$ denotes overdensity parameter in a matter-dominant universe). Thus, the linear evolution of overdensity parameter (Appendix B) can be written as
\begin{equation}
\delta^{\prime\prime}(a)+\Bigg{(}\frac{3}{a}+\frac{E^{\prime}(a)}{E(a)}\Bigg{)} {\delta}^{\prime}(a)-\frac{3}{2}\frac{\Omega_{m_{0}}}{a^5E^2(a)}\delta(a)=0.
\label{22}
\end{equation}
Here, we assume that the collapse is spherical
and homogeneous. We also ignore the effects of rotation and shear. Evidently, a better approximation for the spherical collapse can be obtained if one considers an angular momentum. However, note that we do not see shear effect for a sphere, and the shear tensor equals to zero. Nevertheless, we write an equation of evolution of overdensity parameter in a non-linear regime as
\begin{equation}
\begin{aligned}
\delta^{\prime\prime}(a)+ & \Bigg{(}\frac{3}{a}+\frac{E^\prime(a)}{E(a)}\Bigg{)} \delta^\prime(a)-\frac{4}{3}\frac{1}{(1+\delta)}\delta^\prime(a)^2 \\
-& \frac{3}{2}\frac{\Omega_{m0}}{a^5E(a)^2}(1+\delta)\delta(a)=0.
\end{aligned}
\label{23}
\end{equation}
In the equations mentioned above, $\delta$, $E$, and $a$ denote density contrast, dimensionless Hubble parameter, and the scale factor; respectively. From WMAP and Planck's observations we approximately have $\Omega_m=0.3$. We numerically solve the non-linear growth equations of the perturbations. We imposed the initial conditions in a way to have the growth of non-linear perturbations larger than $10^7$. Note that these initial conditions are necessary for the formation of the large scale structures. In Fig. \ref{fig4} and Fig.\ref{fig5}, we plot linear and non-linear density contrast versus the scale factor and compare the results of the quintessence  model with $\Lambda CDM$ and  coupled tachyon model. Therefore, the results of the quintessence  model are similar to the standard cosmological model.
\begin{figure}
\centering
\includegraphics[width=0.8\linewidth]{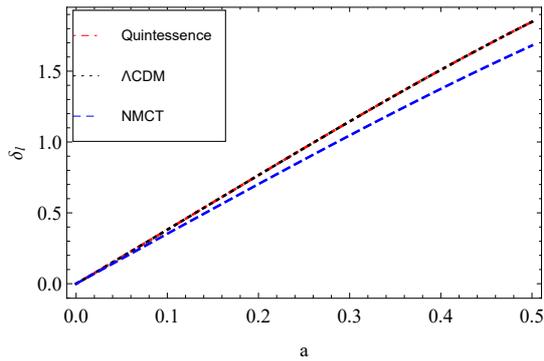}
\caption{The the linear evolution of overdensity parameter $\delta_l$ vs the scale factor for the $\Lambda CDM$, quintessence and NMCT models.}
\label{fig4}
\end{figure}

\begin{figure}
\centering
\includegraphics[width=0.8\linewidth]{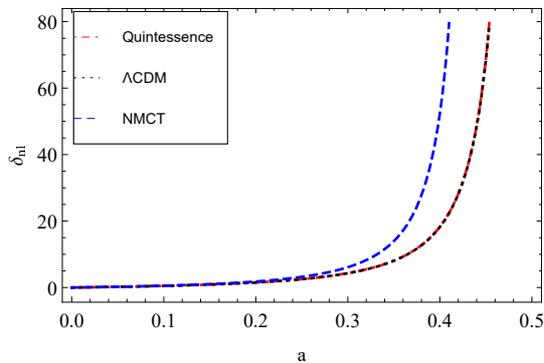}
\caption{The non-linear evolution of overdensity parameter $\delta_{nl}$ versus the scale factor for the $\Lambda CDM$, quintessence and NMCT models.}
\label{fig5}
\end{figure}

\section{Spherical collapse via quintessence field}
\label{sec:2}
\begin{figure}
\centering
\includegraphics[width=0.8\linewidth]{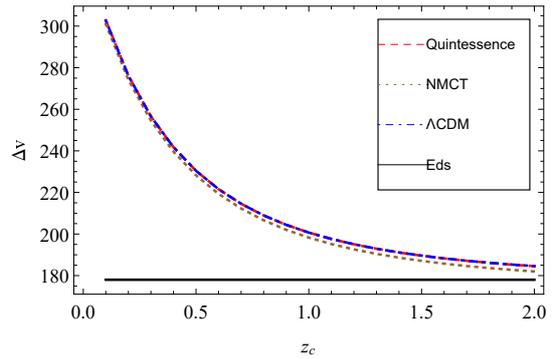}
\caption{The evolution of virialized overdensity parameter $\Delta_V$ versus $z_c$ for the $\Lambda CDM$, quintessence, NMCT and Eds models.}
\label{fig6}
\end{figure}

\begin{figure}
\centering
\includegraphics[width=0.8\linewidth]{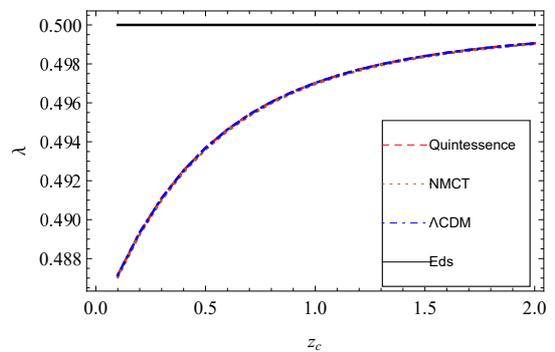}
\caption{The behavior of $\lambda$  versus $z_c$ for the $\Lambda CDM$, quintessence, Eds and NMCT models.}
\label{fig7}
\end{figure}

\begin{figure}
\centering
\includegraphics[width=0.8\linewidth]{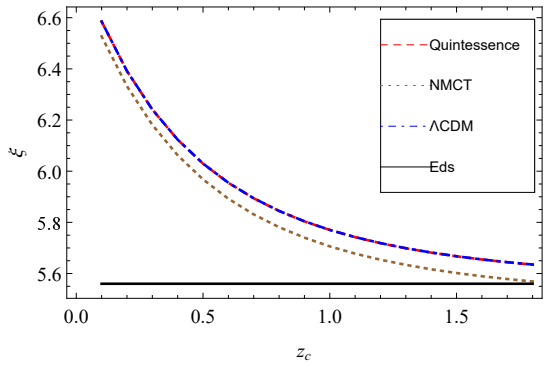}
\caption{The behavior of $\xi$  versus $z_c$ for the $\Lambda CDM$, quintessence, Eds and NMCT models.}
\label{fig8}
\end{figure}

In this section, our goal is to understand how the perturbations in the universe grow within the
framework of the quintessence model and form a dense
structure.  The intended area expands within an expanding background up to a maximum radius $R_{max}$. Then the expansion of the perturbed region stops, and the process of collapse continues. However, it is important to note that the intended area does not collapse to a singular point, and the expanding process ends at half of the maximum radius, which is called virialized radius. Hence, it is essential to realize that the intended area does not collapse to a singular point, and in half of the maximum radius, it is virialized, making the collapse to stop \cite{Maor:2005hq,Meyer:2012nw}. To find the growth of the perturbations, we numerically solve the equation (\ref{22}). As mentioned before, solving this equation is very sensitive to the choice of the initial conditions. Thus, to solve the equation and find the correct value for the linear overdensity parameter $\delta_{c}$. The value of this parameter in the Eds model is constant and equals to 1.68 \cite{Pace:2010sn}. However, one can see from Fig. \ref{fig6} that overdensity parameter changes with the redshift, and eventually, both models share the value of the Eds model. To determine the virialized overdensity parameter, let us consider the sphere with density in the background of $\rho_b$ as defined
below
\begin{equation}
\Delta_V =\frac{\rho}{\rho_b}\frac{R}{a_c},
\label{24}
\end{equation}
in which $R$ is the radius of the structure, $a_c$ denotes the virialized scale factor and $\rho$ is the density of the structure. Equation (\ref{24}) can be rewritten as
\begin{equation}
\Delta_V= 1+\delta (a_c)= \xi (\frac{x_c}{\lambda})^3,
\label{25}
\end{equation}
where $x_c$ is defined as $a_c/a_{ta}$. $a_c$ and $a_{ta}$ denote the scale factor in the virialization and the turn-around epochs; respectively. In the equation (\ref{25}), $\lambda$ and $\xi$ present the radius of structures and perturbations in the turn-around time. In the Eds model, The value of $\Delta_V$ is constant and equals to 178. However, in the other two models, it varies with the redshift. But for the large redshifts, $\Delta_V$ has the same value as the Eds model. In Fig .\ref{fig6}, we plot the virialized overdensity parameter versus $z_{c}$ for the quintessence, $\Lambda CDM$, NMCT, and Eds models. Virialized overdensity is a constant for the Eds model, and it equals to 178. However, for the other models, it varies with the redshift. For the quintessence model and the standard model, the value of this parameter is approximately the same. But in the tachyonic model, the slope of the graph approaches the Eds model slightly faster, which means that the perturbations in the tachyonic model turn to virialized overdensity faster. Thus, the possibility for the formation of the structures in this model occurs sooner than the other two models.\\
As we mentioned earlier, the growth of the perturbations during the collapse phase is virilized at half of the maximum radius, and so the structures can be formed. At the turn-around moment, we plot the radius of the cosmological structures $\lambda $ for the four models, as shown in Fig. \ref{fig7}. One can see that $\lambda $ takes its maximum value for the Eds model, which is $0.5$. However, in the other models, the value of $\lambda$ depends on the redshift and for higher redshifts, it tends to the Eds values.\\
The $\xi$ parameter that appears in equation (\ref{25}) is as follows
\begin{equation}
\xi = \frac{\rho(R_{ta})}{\rho_b}=1+\delta(a_{ta}).
\label{26}
\end{equation}
It can be seen from Fig. \ref{fig8} that the parameter $\xi$ for the Eds model is a constant and equal to 5.6. But, for the other models, it varies with respect to virialized redshift, and for large redshifts takes a value which is close to what it has in the Eds model. But the coupled tachyonic model tends to the Eds model with a steeper slope, which means that within the framework of the tachyonic model, during virialization, the structures take the return path faster than other models. Therefore, it can be concluded that in this model the structures are formed faster than the other models.\\

\section{Tachyon field with a constant potential}
\label{sec:3}
\begin{figure}
\centering
\includegraphics[width=0.8\linewidth]{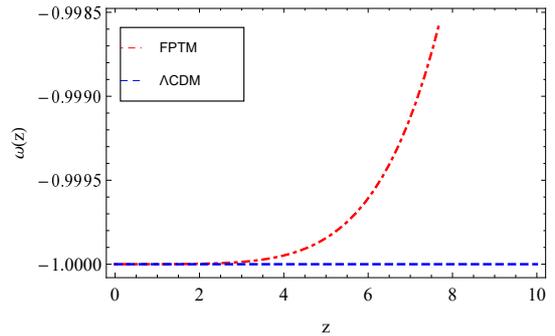}
\caption{The behaviour of Eos parameter $\omega$  versus $z$ for the $\Lambda CDM$ and FPTM models.}
\label{fig9}
\end{figure}

\begin{figure}
\centering
\includegraphics[width=0.8\linewidth]{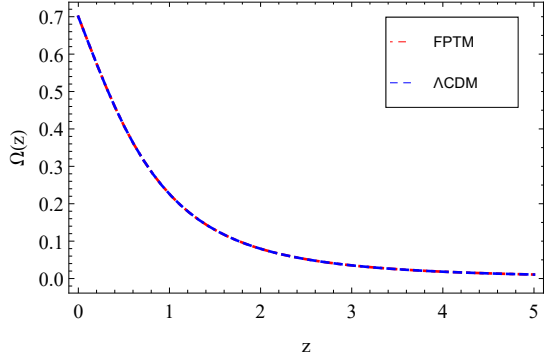}
\caption{The variation of energy density parameter $\Omega(z)$  versus $z$ for the $\Lambda CDM$ and FPTM models.}
\label{fig10}
\end{figure}

\begin{figure}
\centering
\includegraphics[width=0.8\linewidth]{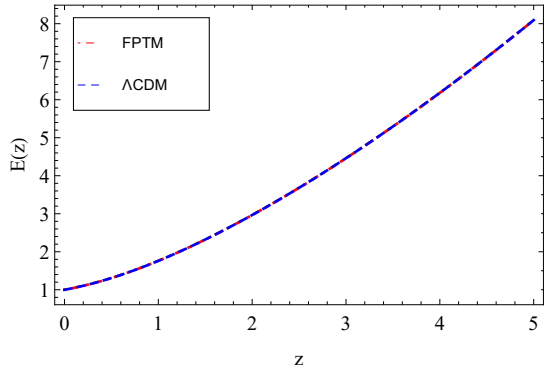}
\caption{The variation of the Hubble dimensionless parameter $E(z)$ versus $z$ for the $\Lambda CDM$ and FPTM models.}
\label{fig11}
\end{figure}

\begin{figure}
\centering
\includegraphics[width=0.8\linewidth]{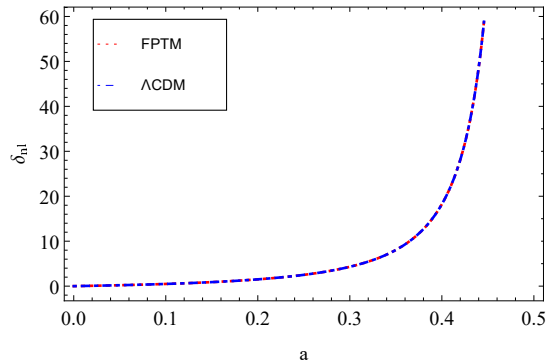}
\caption{The behaviour of the non-linear tachyonic perturbations $\delta_l$ versus $a$ for the $\Lambda CDM$ and FPTM models.}
\label{fig12}
\end{figure}

\begin{figure}
\centering
\includegraphics[width=0.8\linewidth]{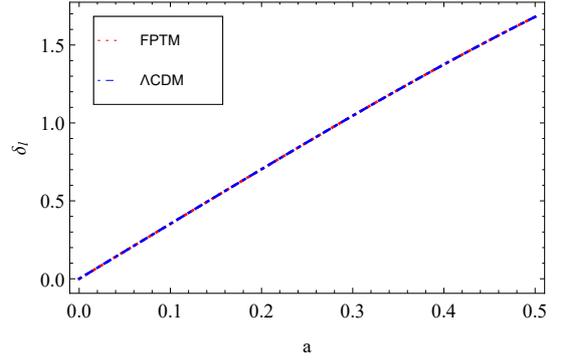}
\caption{The behaviour of the linear tachyonic perturbations $\delta_l$ versus $a$ for the $\Lambda CDM$ and FPTM models.}
\label{fig13}
\end{figure}

\begin{figure}
\centering
\includegraphics[width=0.8\linewidth]{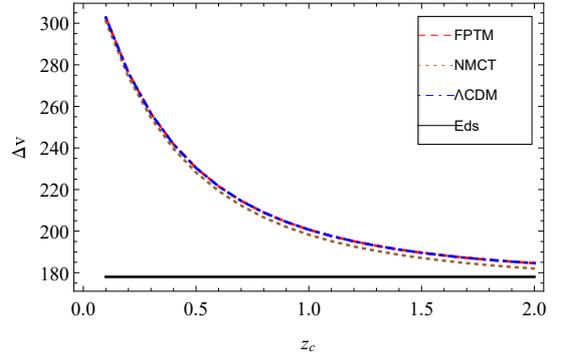}
\caption{The evolution of virialized overdensity parameter $\Delta_V$ versus $z_c$ for the $\Lambda CDM$, FPTM, NMCT and Eds models.}
\label{fig14}
\end{figure}

\begin{figure}
\centering
\includegraphics[width=0.8\linewidth]{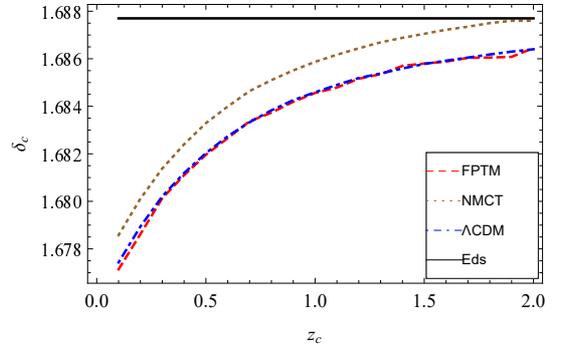}
\caption{The behaviour of the linear overdensity parameter $\delta_c$  versus $z_c$ for the $\Lambda CDM$, FPTM, NMCT and Eds models.}
\label{fig15}
\end{figure}

\begin{figure}
\centering
\includegraphics[width=0.8\linewidth]{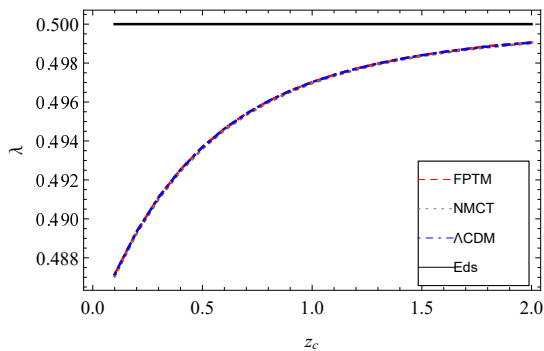}
\caption{The evolution of the radius of structure $\lambda$ versus $z_c$ for the $\Lambda CDM$, FPTM, NMCT and Eds models.}
\label{fig16}
\end{figure}

\begin{figure}
\centering
\includegraphics[width=0.8\linewidth]{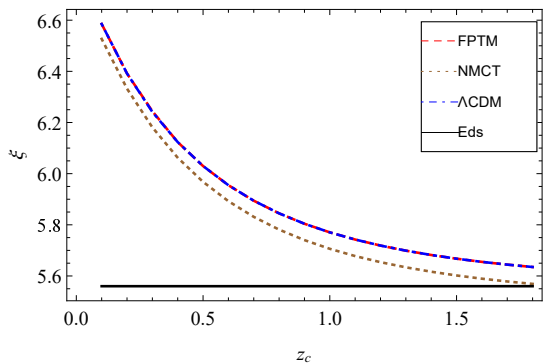}
\caption{The evolution of the $\xi$ parameter versus $z_c$ for the $\Lambda CDM$, FPTM, NMCT and Eds models.}
\label{fig17}
\end{figure}

Let us assume that the Chaplygin gas acts as a tachyon field, so its Lagrangian becomes
\begin{equation}
L=-V(\phi)\sqrt{1-\dot{\phi}^2}.
\end{equation}
The potential in this model is a constant and equals to $V_0$. Using the Euler-Lagrange method, one can obtain the equation of motion as
\begin{equation}
\frac{\ddot{\phi}}{1-\dot{\phi}^2}+3H\dot{\phi}+\frac{1}{V(\phi)}\frac{dV(\phi)}{d\phi}=0.
\end{equation}
Using the energy-momentum tensor, we find the energy density and pressure for the tachyon field
\begin{equation}
T_{0}^0=\rho_{\phi}=\frac{V(\phi)}{\sqrt{1-\dot{\phi}}},
\end{equation}
\begin{equation}
T_{i}^i =P_\phi=-V(\phi)\sqrt{1-\dot{\phi}^2}.
\end{equation}
The EoS parameter corresponding to the tachyon field is
\begin{equation}
\omega_\phi =\frac{p_{\phi}}{\rho_\phi}= \dot{\phi}^2-1.
\end{equation}
So, the energy density of the tachyon and matter field are as follows
\begin{equation}
\rho_\phi=\rho_{\phi0}a^{-3(1+\omega_\phi)} \:\: and \:\: \rho_m=\rho_{m0}a^{-3},
\end{equation}
with the continuity equations
\begin{equation}
\dot{\rho_\phi}+3H\rho_\phi \big{(}1+\omega_\phi\big{)}=0,
\end{equation}
and
\begin{equation}
\dot{\rho_m}+3H\rho_m=0.
\end{equation}
We know that the tachyon energy density is
\begin{equation}
\dot{\Omega}_\phi=-\Omega_\phi H\big{[}3(1+\omega_\phi)+2\frac{\dot{H}}{H}\big{]}.
\label{37}
\end{equation}
Using the Friedmann and the continuity equations, we find the expression for
\begin{equation}
\frac{\dot{H}}{H}=-\frac{3}{2}(1+\omega_\phi \Omega_\phi).
\end{equation}
Using the change of variable $\frac{d}{d(\ln a)}= -(1+z)\frac{d}{dz}$ , the equation (\ref{37}) can be rewritten as
\begin{equation}
\Omega^\prime_\phi=-3\Omega_\phi\omega_\phi(\Omega_\phi-1)(1+z)^{-1},
\end{equation}
where prime represents the derivative with respect to $z$. According to the Planck's recent results, this parameter is approximately $0.7$, but tends to zero for the very large redshifts \cite{Pace:2010sn}. As discussed in the previous section, the spherical collapse model for a fixed-potential tachyonic model (FPTM) which is equivalent to a Chaplygin gas can be studied. In Fig. \ref{fig9}, the EoS parameter $\omega$ is plotted for both the FPTM and the $\Lambda CDM$ models. One can see that, for the small redshifts, the value of $\omega_\phi$ approaches to $-1$. In Fig. \ref{fig10}, the behaviour of energy density of the FPTM is compared with the standard model of cosmology. The plot shows that the variation of $\Omega(z)$ for the different redshifts is similar for both of the models.\\
 The Hubble dimensionless parameter $E(z)$ is plotted in Fig. \ref{fig11}. This graph shows the compatibility of the fixed-potential tachyon model with the standard cosmological model.
We solve equations (\ref{22}) and (\ref{23}) for the FPTM and obtained the growth of linear and non-linear tachyonic perturbations. In Fig. \ref{fig12} and Fig. \ref{fig13}, we plot the growth of linear and non-linear perturbations in terms of a function of the scale factor.
Then to describe the spherical collapse model in the framework of the fixed-potential tachyonic model, we need to solve equations (\ref{25}) and (\ref{26}) for the FPTM state. We numerically solved these equations and plot the evolution of virialized overdensity $\Delta_V$, linear overdensity $\delta_c$, radius of the structure $\lambda$ and $\xi$ parameters in figures \ref{fig14}, \ref{fig15}, \ref{fig16} and \ref{fig17}; respectively. These graphs show that in the tachyonic model, the $\delta_c$ and $\xi$ curves tend to Eds model faster than the quintessence and standard models. Therefore, we conclude that within the framework of the tachyon model, the perturbations are virialized faster, and the structures are formed earlier.

\section{Conclusion}
\label{sec:4}
In this paper, we used the Chaplygin gas model to describe the spherical collapse of the large-scale structures. we solved the quintessence model with potential $ V(\phi)=\sqrt{A} (\cosh^2 (\sqrt{3}\phi)+1)/(2\cosh(\sqrt{3}\phi))$. Numerical analysis of this model showed that for $0.28 \leq A \leq 0.55$, our model agrees with the current observational results. Then, we obtained the equation of state parameter for this model. As we expected, the EoS parameter of our model changed over time, and in the case of small redshifts, the standard cosmological model was favored. Also, using the parameter of the equation of state and the continuity equation, we obtained the evolution of the Hubble dimensionless parameter and the energy density of the scalar field. The results showed that  within the framework of the Chaplygin gas model, during virialization, the structures take the return path faster
than other models. Therefore, it can be concluded that in
this model, the structures are formed faster than the other
models. Then, using the numerical solution of the differential equations of growth of non-linear and linear perturbation, we investigated the structure formation parameters. In studying the structure formation of the spherical collapse model, we obtained parameters such as structure radius, virialized overdensity $\Delta_V$ and $\xi$ parameters.
By considering the Chaplygin gas as an ordinary tachyonic scalar field, $\phi$, one can obtain a corresponding potential for it. Finally, we considered the fixed-potential tachyon model as another candidate for the Chaplygin gas and calculated all the parameters of the spherical collapse for it. During virialization, the results confirmed that in the framework of the fixed potential tachyonic model, the structures take the return path faster
than other models. Therefore, it can be concluded that in
this model, the structures are formed faster than the other
models. Therefore, we conclude that the Chaplygin gas model can properly explain the spherical collapse of the large-scale structures.%
\section{Appendix A: Quintessence and tachyon equivalence to the Chaplygin gas }
The simplest form of the equation of state for the Chaplygin gas is
\begin{equation}
P=-\frac{A}{\rho},\label{388}
\end{equation}
where $A$ is a positive constant. Equation (\ref{388}) is a particular form for the tachyon's equation of state with a constant potential. The general form of the equation of state for the tachyonic model is
\begin{equation}
P=-\frac{V^2(\phi)}{\rho},\label{39}
\end{equation}
where $\phi$ denotes the tachyonic field. In equation (\ref{39}), one can treat the potential as a constant, and find the EoS for the Chaplygin gas. Now, using equation (\ref{388}) and the continuity equation $\dot{\rho}+3(\dot{a}/a)(\rho+P)=0$, we find
\begin{equation}
\rho=\sqrt{A+\frac{B}{a^6}},\label{40}
\end{equation}
where A and B are positive constants and $a$ is the scale factor. One can see that for $a<<(B/A)^{1/6}$, the energy density $\rho$ will be in the order of $\sim \sqrt{B}/a^3$ and when $a$ is much larger than $(B/A)^{1/6}$, $\rho$ tends to $-P\simeq \sqrt{A}$. Thus, one can conclude that in the past, when the scale factor was smaller, the Chaplygin gas can be considered as a pressureless matter. As we get closer to the present time, the scale factor grows, and the density of the Chaplygin gas becomes the same as the density of cosmological constant. Therefore, this fluid can be the source of undergoing accelerated expansion at the current time. Now, we will show that by choosing a particular potential in the quintessence model \cite{Copeland:2006wr}, we obtain the equation of state of the Chaplygin gas. In the quintessence model, the energy density of the scalar field and its pressure are written as $\rho=\dot{\phi}^2/2+V(\phi)$ and $P=\dot{\phi}^2/2-V(\phi)$. Using these equations, we find
 \begin{equation}
\dot{\phi}^2=P+\rho\label{41},
\end{equation}
\begin{equation}
V(\phi)=(\rho-P)/2.
\end{equation}
Using equations (\ref{388}) and (\ref{40}) we have
\begin{equation}
\dot{\phi}^2=\frac{B}{a^6\sqrt{A+B/a^6}},\label{43}
\end{equation}
\begin{equation}
V(\phi)=\frac{1}{2}[a^6\sqrt{A+B/a^6}+\frac{A}{a^6\sqrt{A+B/a^6}}]\label{44}.
\end{equation}
Applying the change of variable $\frac{\partial}{\partial t}=-aH\frac{\partial}{\partial a}$ and using $H^2=(8\pi G/3) \rho$, equation (\ref{43}) becomes
\begin{equation}
\frac{\kappa}{\sqrt{3}}\frac{d\phi}{da}=\frac{\sqrt{B}}{a\sqrt{Aa^6+B}},
\end{equation}
where $\kappa$ equals to $\sqrt{8\pi G}$. Integrating the above equation, we obtain
\begin{equation}
a^6= \frac{4Be^{2\sqrt{3}\kappa \phi}}{A(1-e^{2\sqrt{3}\kappa \phi})^2},\label{46}
\end{equation}
Substituting equation (\ref{46}) into equation (\ref{44}), we find the proper form of the potential function as follows
\begin{equation}
V(\phi)=\frac{\sqrt{A}}{2}\bigg{(} \cosh(\sqrt{3}\kappa\phi)+\frac{1}{\cosh(\sqrt{3}\kappa\phi)}\bigg{)}.
\end{equation}

\section{Appendix B: The evolution of overdensity parameter}
To investigate the cosmic fluid, we use the continuity, Euler, and Poisson equations, which describe the dynamic of the fluid. These equations are\begin{equation}
\frac{\partial{\rho}}{\partial{t}}+\nabla_r .(\rho v)+p \nabla_r .v=0,
\end{equation}
\begin{equation}
\frac{\partial{\rho}}{\partial{t}}+v.\nabla_r .(v )+p \nabla_r \phi+\frac{\nabla_rP+v\dot{P}}{\rho+P}=0,
\end{equation}
\begin{equation}
\nabla^2\phi=4\pi G (\rho+3P),
\end{equation}
and
\begin{equation}
\dot{\bar{\rho}}+3H(P+\bar{\rho})=0.
\end{equation}
where $v$ is velocity, and $ \phi$ is gravitational potential. The physical coordinate $\vec{r}$ that is related to the comoving coordinate $\vec{x}$ via $\vec{r}=a\vec{x}$ in which $a$ is the scale factor. Since it is not possible to find a general solution to the non-linear equations, we write the quantities such as density, pressure, and gravitational potential as two parts of the background field and the perturbations around it. Writing these quantities in the comoving coordinates, the energy density, the equation of state, velocity, and the gravitational potential take the following forms
\begin{equation}
\rho(\vec{x},t)=\bar{\rho}[1+\delta(\vec{x},t)],
\end{equation}
\begin{equation}
P=\omega\rho(\vec{x},t),
\end{equation}
\begin{equation}
\vec{v}(\vec{x},t)=a[H(a)\vec{x}+\vec{u}(\vec{x},t)],
\end{equation}
and
\begin{equation}
\Phi=\Phi_0(\vec{x},t)+\phi(\vec{x},t)
\end{equation}
Using the quantities defined above, the equations of the dynamics of the fluid can be written as
\begin{equation}
\dot{\delta}+(1+\omega)(1+\delta)\nabla_{\vec{x}}.u=0,\label{56}
\end{equation}
\begin{equation}
\frac{\partial{\vec{u}}}{\partial{t}}+2H\vec{u}+(\vec{u}.\nabla_{\vec{x}})\vec{u}+\frac{1}{a^2}\nabla_{\vec{x}}\phi=0\label{57}
\end{equation}
and
\begin{equation}
\nabla^2\phi-4\pi G(1+3\omega)a^2\bar{\rho}\delta=0.\label{58}
\end{equation}
Taking the divergence from (\ref{57}), we have
\begin{equation}
\nabla.[(\vec{u}.\nabla)\vec{u}]=\frac{1}{3}\theta^2+\sigma^2-\omega^2,\label{59}
\end{equation}
where $\theta=\nabla_{\vec{x}}.\vec{u}$. Also, the stress tensor $\sigma_{ij}$ and the rotational tensor $\omega_{ij}$ are defined as
\begin{equation}
\sigma_{ij}=\frac{1}{2}(\frac{\partial{u^j}}{\partial{x^i}}+\frac{\partial{u^i}}{\partial{x^j}})-\frac{1}{3}\theta \sigma_{ij},
\end{equation}
\begin{equation}
\omega_{ij}=\frac{1}{2}(\frac{\partial{u^j}}{\partial{x^i}}-\frac{\partial{u^i}}{\partial{x^j}})
\end{equation}
which can be used to define $\omega^2=\omega_{ij}\omega^{ij}$ and $\sigma^2=\sigma_{ij}\sigma^{ij}$. Now if we take derivative from equation (\ref{56}) with respect to time and use equations (\ref{56}, \ref{57}, \ref{58} and \ref{59}) , we obtain the following equation for the evolution of the spherical overdensity
\small{\begin{equation}
\begin{split}
\ddot{\delta}&+(2H-\frac{\dot{\omega}}{1+\omega})\dot{\delta}-\frac{4+3\omega}{3(1+\omega})\frac{\dot{\delta}^2}{1+\delta}-\\
&4 \pi G \bar{\rho}(1+\omega)(1+3\omega)\delta(1+\delta)-(1+\omega)(1+\delta)(\sigma^2-\omega^2)=0.\label{62}
\end{split}
\end{equation}}
Using the change of variable $\frac{\partial}{\partial t}=-\dot{a}\frac{\partial}{\partial a}$, we rewrite equation (\ref{62}) as follows
\small{\begin{equation}
\begin{split}
\delta^{\prime\prime}(a)&+(\frac{3}{a}+\frac{E^{\prime}(a)}{E(a)}-\frac{\omega^{\prime}}{1+\omega})\delta^\prime(a)\\&-\frac{4+3\omega}{3(1+1+\omega)}\frac{\delta^{\prime^2}(a)}{1+\delta(a)}\\
& \frac{3}{2}\frac{\Omega_{m_0}}{a^2E^2(a)}h(a)(1+\omega)(1+3\omega)\delta(a)(1+\delta(a))\\&-\frac{1}{aH^2(a)}(1+\omega)(1+\delta(a))(\sigma^2-\omega^2)=0,
\end{split}
\end{equation}}
where $E(a)=H/H_0$ and $\omega_{m0}$ denotes the density parameter of matter in present. In the spherical collapse model, we can ignore the stress and rotational tensors. Also, we assume the matter is pressureless. So for the nonlinear regime, we obtain
\begin{equation}
\begin{split}
\delta^{\prime \prime}+\bigg{(}\frac{3}{a}+&\frac{E^\prime(a)}{E(a)}\bigg{)}\delta^\prime-\frac{4}{3}\bigg{(}\frac{1}{1+\delta}\bigg{)}\delta^{\prime^2}-\\
\frac{3}{2}\bigg{(}&\frac{\Omega_{m_0}}{a^5E^2}\bigg{)}\delta(1+\delta)=0.
\end{split}
\end{equation}
For the linear regime, the equation for the evolution of the spherical overdensity becomes
\begin{equation}
\delta^{\prime \prime}+\bigg{(}\frac{3}{a}+\frac{E^\prime(a)}{E(a)}\bigg{)}\delta^\prime-\frac{3}{2}\bigg{(}\frac{\Omega_{m_0}}{a^5E(a)^2}\bigg{)}\delta=0
\end{equation}

%

\end{document}